\documentstyle[aps,manuscript,epsf]{revtex}

\def\be{\begin{equation}}
\def\ee{\end{equation}}
\def\bea{\begin{eqnarray}}
\def\eea{\end{eqnarray}}
\newcommand{\frat}[2]{\frac{\textstyle #1}{\textstyle #2}}
\newcommand{\vf}[1]{\mbox{\boldmath $#1$}}

\begin{document}

\title{ \rightline{\small KANAZAWA 99-04}
       Field Strength Correlators and Gluon Condensates  \\
       at Finite Temperature from Renormalization Group Smoothing}

\author{
    Ernst-Michael Ilgenfritz$^{1}$ and Stefan Thurner$^{2}$\\
    $^{1}${\it Institute for Theoretical Physics, Kanzawa University, Japan} \\
    $^{2}${\it Institut f\"ur Kernphysik, TU-Wien, Austria }\\
}

\maketitle
\begin{abstract}
We summarize recent attempts to extract characteristics of non-perturbative
vacuum structure from lattice measurements of the gauge invariant field 
strength correlator. As an alternative to cooling, we propose to apply the 
renormalization group (RG) smoothing method in lattice studies. 
For pure $SU(2)$ gauge theory we present magnetic and electric
correlation lengths and condensates related to various correlators
over a temperature range of 
$0.7~T_{\mathrm{dec}} < T < 1.9~T_{\mathrm{dec}}$.
\end{abstract}

\newpage 

\section{Introduction}

Quantum Chromodynamics (QCD) owes its confining property to  
non-perturbative structures of its groundstate which are, 
after many years of research, not yet uniquely identified.
(for a recent overview of competing pictures see e.g. \cite{Lattice98}).
In a more pragmatic attitude, the Stochastic Vacuum Model (SVM) \cite{SVM}  
describes the vacuum structure in terms of a few vacuum correlation
functions and points out which are related to confinement.
That only few correlators should characterize the vacuum is the consequence
of a truncation of the hierarchy of correlation functions, essentially due 
to a postulate made within the SVM that the gauge fields are 
Gaussian random processes in Euclidean space-time.
In this framework, the correlators enter,  and relate to each 
other,  various phenomenological predictions. Particularly interesting 
(exhausting, in the case of pure gluodynamics) is the gauge invariant 
gluon field strength correlator in Euclidean space (and its continuation 
to Minkowski space). According to the SVM, part of the Euclidean field 
strength correlator at zero temperature is related to the confinement 
property of the QCD ground state \cite{Confinement}.  
For heavy quarkonium states, the full correlator describes the effect 
of the gluon condensate on the level splitting in 
detail (see Ref. \cite{Quarkonia} and references therein). Wick-rotated to 
Minkowski space, the correlator serves to parametrize high energy 
hadron-hadron scattering (see e.g. Ref. \cite{Scattering}). 

Regardless of how this truncation exactly works, these correlators 
incorporate the unknown non-perturbative mechanisms and can be 
considered as an implicit parametrization of the important gluonic vacuum 
field modes.  If they were  known, this information could  be  readily 
used to constrain the parameters of more explicit models of vacuum structure.
The central role, both theoretically and phenomenologically,
strongly suggests to extract the field strength correlator 
(and other vacuum correlators) from a first principle lattice simulation. 
It is interesting to notice that, although the SVM approach has been 
proposed more than a decade ago by Dosch and Simonov \cite{SVM}, 
on the lattice 
the field strength correlator has only been studied by the Pisa group. 
Apart from earlier results for $SU(2)$ \cite{Campostrini} 
and $SU(3)$ \cite{DiGiacomo1}
pure gauge theory, 
there are now precise results available, from an analysis using cooling, 
over an interval of distances 0.1 ... 1.0 fm for pure 
$SU(3)$ gluodynamics \cite{DiGiacomo2,DiGiacomo3}, 
for QCD with dynamical (Kogut-Susskind) fermions 
\cite{DiGiacomo4} (four flavors, for two values of mass) 
over an interval 0.4 ... 1.0 fm,  and, over the same interval of distances, 
for pure $SU(3)$ Yang-Mills theory very near the deconfining 
temperature \cite{DiGiacomo3}. 

Lattice data for $T=0$ obtained and fitted in Refs.
\cite{DiGiacomo2,DiGiacomo3,DiGiacomo4} give strong hints in favor of 
dominance of locally selfdual or antiselfdual field strengths.
These data have been analyzed \cite{IMMMS} from the point of view of the 
random instanton liquid model \cite{RILM,Shuryak-review}. 
Here the following picture arises: a rather dilute liquid of instantons 
(with parameters which, in the case of full QCD, happen to be roughly 
consistent with parameters obtained from phenomenological analyses) can 
explain the field strength correlator, including the part which is believed 
to be responsible for confinement. The relevance of this statement
for confinement is far from clear. Absolute confinement is, strictly 
speaking, a property of pure Yang-Mills theory alone. 
At this place let us mention that a much higher instanton density is
extracted from the field strength correlator in the case of pure Yang-Mills 
theory than for full QCD \cite{IMMMS}.

On the other hand, according to a widespread opinion a random instanton 
liquid, however dense it might be, does not lead to an asymptotic string 
tension. This is obvious for calculations of the direct effect of the 
(unperturbed) instanton liquid on the static $Q$-$\bar{Q}$ potential via
the Wilson loop. While the instanton potential rises linearly at intermediate 
distance, this does not correspond to the known string tension \cite{Chen}
if the phenomenological density and size are assumed to hold. 
Let us note, however, that neither 
instanton suppression by external charges or fields nor instanton interactions 
(see {\it e.g.} \cite{instanton_orient}) have been taken into account in these 
estimates. It has even been questioned \cite{Diakonov_string} whether the 
available lattice evidence for confinement is conclusive at all.
In short, in spite of the possibility of an instanton interpretation 
of the correlator, 
the role of instantons for confinement is still controversial. 
The answer depends on details of the modeling of the instanton liquid, in
particular on the density and the shape of size distribution 
(see  e.g. Refs. \cite{Boulder_cycling,Suganuma}). 

There is an indirect route (not completely worked out yet) leading from the 
instanton liquid to confinement. It has been demonstrated that under certain 
circumstances an instanton liquid induces the condensation (percolation)
of magnetic monopoles (see Ref.\cite{Suganuma} and earlier references 
therein) which should confine via a dual Meissner effect.   
A corresponding effective theory of a dual Abelian gauge field, interacting 
minimally with the (condensed) monopole field (Dual Abelian Higgs Model
\cite{DAHM}) is characterized by the mass of the dual ``photon'' and a 
finite Higgs correlation length. Both quantities together determine 
\cite{Baker} the gauge invariant non-Abelian field strength correlator which 
we are interested in. Therefore, in a more fundamental way, the two structure 
functions of the correlator at $T=0$ are inherent to an infrared 
effective dual formulation of QCD (see also \cite{Ellwanger}).  

Similarly to $T=0$ the field strength correlator at finite temperature encodes 
information on generic non-perturbative structures of non-Abelian gau\-ge 
field histories dominating the QCD partition function. 
In a similar spirit as of Ref. \cite{IMMMS}, the correlators obtained near 
the deconfinement transition in pure $SU(3)$ gauge theory 
\cite{DiGiacomo2,DiGiacomo3} have been analyzed from the caloron (finite $T$ 
instanton) point of view \cite{IMM}.  Such a description turns out to be 
possible only below the deconfinement temperature $T_{\mathrm{dec}}$.

As any measurement involving the field strength, such lattice calculations 
are far from trivial. The cooling method has been used in most of the more
recent studies. To our opinion, the implications of this method for the 
intended measurements have not been discussed thoroughly enough so far.  
In this paper, we are going to apply a different technique to improve the 
signal for extracting the field strength correlator from lattice gauge field
ensembles. 
Our method is based on a (classically) perfect action and uses a 
renormalization group motivated smoothing procedure \cite{Feurstein,PRD98}.
We shall give a presentation of first results obtained in this manner for 
finite temperature $SU(2)$ pure gauge theory.
In contrast to $T=0$, a similar splitting of correlation lengths related 
to the different structure functions is obtained with both techniques.
Some possible differences should be closer examined whether they
are due to the other gauge group $SU(2)$ used here, compared to
$SU(3)$ analyzed with cooling. If not then, cooling would 
have a systematic effect concerning the relative strength of the structure 
functions ${\cal D}$ and ${\cal D}_1$ at high temperature.

The paper is organized as follows. In Sec. II we give an introduction to 
the concept of field strength correlators for zero and finite temperature; 
the correlators and their suitable decompositions into the structure functions
are defined there. In Sec. III we briefly discuss the potential problems 
of the cooling method and recall in short the fixed point action 
renormalization group (RG) smoothing technique. In Sec. IV we describe 
our results, and conclude with a discussion in Sec. V. 

\section{Field Strength Correlator}

The two-point field strength correlator is a non-local generalization of 
the gluon condensate, involving the field strength tensor at separate points :
\be\label{eq:basic_def}
{\cal D}_{\mu\rho\nu\sigma}(x) = g^2 \langle 0|
\mathrm{Tr} \left\{ G_{\mu\rho}(0) S(0,x) G_{\nu\sigma}(x) S^\dagger(0,x) \right\}
|0 \rangle ~,
\ee
where $G_{\mu\rho} = T^a G^a_{\mu\rho}$ 
is the field strength,  with  $T^a$ being the generators of the
algebra of the color $SU(N_c)$ group in the fundamental representation.
$S(x_1,x_2)$ is the (path ordered) Schwinger line (parallel transporter)
\be\label{eq:schwinger_line}
S(x_1,x_2) \equiv {\rm P} \exp \left( i g \int_{x_1}^{x_2} 
dz^\mu A_\mu (z) \right) ~,
\ee
with $A_\mu = T^a A^a_\mu$.
It is necessary to include the non-Abelian Schwinger line phase factor 
in order to ensure gauge invariance. Usually 
the straight--line path from $x_1$ to $x_2$ is chosen to evaluate $S(x_1,x_2)$.
The correlator describes the coherence of the gauge field 
(modulo gauge transformations) over some coherence length. 
This notion replaces the idea of ``domains'' originally imagined to form the 
gluon condensate. Completely contracted, it is the local gluon condensate
defined as
\be\label{eq:gluon_cond}
G_2 \equiv \langle {\alpha_s \over \pi} :G^a_{\mu\nu} G^a_{\mu\nu}: \rangle
~,~~~~~~~ (\alpha_s = {g^2 \over 4\pi}) ~,
\ee
which is obtained in the limit $|x_1-x_2| \rightarrow 0$ from the regulated
correlator. 
	
\subsection{Zero Temperature}	
	
At $T=0$, Euclidean $O(4)$ invariance admits the decomposition (for $x=x_1-x_2$)
\bea\label{eq:decomp_T0}
\lefteqn{
{\cal D}_{\mu\rho\nu\sigma}(x) = (\delta_{\mu\nu}\delta_{\rho\sigma} -
\delta_{\mu\sigma}\delta_{\rho\nu})
\left[ {\cal D}(x) + {\cal D}_1(x) \right] } \nonumber \\
& & + (x_\mu x_\nu \delta_{\rho\sigma} - x_\mu x_\sigma \delta_{\rho\nu}
+ x_\rho x_\sigma \delta_{\mu\nu} - x_\rho x_\nu \delta_{\mu\sigma})
{\partial{\cal D}_1(x) \over \partial x^2} ~,
\eea
where ${\cal D}$ and ${\cal D}_1$ are two invariant functions (of 
Euclidean modulus $|x|$) containing information on the vacuum.
The tensor structure related to ${\cal D}_1$ does not contribute to confinement,
while ${\cal D}\ne0$ is possible only in non-Abelian theories or in 
Abelian theories with monopoles. 

It is convenient, in particular on the lattice,
to define two other vacuum structure functions
${\cal D}_\parallel(x)$ and a
${\cal D}_\perp(x)$, according to whether the plaquettes defining
the field strength component in question have one direction parallel to the
4-vector between the two points of observation.
These functions are composed of ${\cal D}$  and ${\cal D}_1$ as follows
\bea\label{eq:par_perp}
{\cal D}_\parallel &\equiv& {\cal D} + {\cal D}_1 + x^2 {\partial{\cal D}_1
\over \partial x^2} ~, \nonumber \\
{\cal D}_\perp &\equiv& {\cal D} + {\cal D}_1 ~.
\eea
The two basic vacuum structure functions at $T=0$ have a non-perturbative
and a perturbative part according to the notation
\bea\label{eq:nonpert_pert}
{\cal D}(x) &=& {\cal D}^{np}(x) 
+ {a_0 \over |x|^4} {\rm e}^{-|x|/\lambda_0} ~, \nonumber \\
{\cal D}_1(x) &=& {\cal D}_1^{np}(x) 
+ {a_1 \over |x|^4} {\rm e}^{-|x|/\lambda_1} ~,
\eea
Since the first structure function is closely related to confinement, 
the non-perturbative parts of the two functions are suggestively written as
\bea\label{eq:non_pert}
{\cal D}^{np}(x) &=& {\pi^2 \over 6} \kappa G_2 \tilde{\cal D}(x)
 ~, \nonumber \\
{\cal D}_1^{np}(x) &=& {\pi^2 \over 6} (1-\kappa) G_2 \tilde{\cal D}_1(x)
 ~,
\eea
where $G_2$ has to be identified with the gluon condensate 
(\ref{eq:gluon_cond}) and 
\be\label{eq:normaliz}
 \tilde{\cal D}(0) = \tilde{\cal D}_1(0) = 1 \, . 
\ee
The so-called non-Abelianicity
$\kappa={\cal D}^{np}(0)/\left({\cal D}^{np}(0)+{\cal D}_1^{np}(0)\right)$ 
represents a certain fraction of the gluon condensate related to confinement.
So far the functions $\tilde{\cal D}(x)$ and $\tilde{\cal D}_1(x)$ 
mostly have been fitted as exponential functions 
\cite{DiGiacomo1,DiGiacomo2,DiGiacomo3,DiGiacomo4}
of the modulus $|x|$
\be\label{eq:shape}
\tilde{\cal D}(x) = {\rm e}^{-|x|/\Lambda_0} \quad 
{\rm and} \quad 
\tilde{\cal D}_1(x) = {\rm e}^{-|x|/\Lambda_1} ~,
\ee
although more physically motivated forms are available which 
fit the data as well \cite{Meggiolaro}.

In the SVM the string tension can be approximately related to the 
correlation length $\Lambda_0$ (coherence length of ${\cal D}$) 
and the gluon condensate \cite{SVM},
\begin{eqnarray}\label{eq:sigma}
\sigma&=& \frac{1}{2} \int d^2x {\cal D}(x) 
\nonumber\\ [-.2cm]
\\ [-.25cm]
&=& \frac{\pi^3}{12}~\kappa~G_2 \int dx^2 \tilde{\cal D}(x) 
= \frac{\pi^3}{6}~\kappa~G_2~\Lambda_0^2 \,  . 
\nonumber
\end{eqnarray}

In the perturbative regime (at short distance) both ${\cal D}$ and 
${\cal D}_1$  behave as $1/|x|^4$. 
One-gluon exchange contributes only to the latter \cite{Shevchenko1} 
giving ${\cal D}_1=\frac{16~\alpha_s}{3~\pi~|x|^4}$.
One-loop and higher order perturbative contributions to the string tension
integral (\ref{eq:sigma})
are cancelled by higher correlator contributions \cite{Shevchenko2}.

Purely selfdual or antiselfdual non-perturbative background fields  
contribute only to ${\cal D}$ (see Ref. \cite{Dorokhov97,IMMMS}). 
In this case, it is the overlap of different selfdual or antiselfdual
configurations
(higher order contributions in an expansion in the density \cite{IMMMS}) 
which generates the non-perturbative part of ${\cal D}_1$.
The observation that ${\cal D}$ $ >>$ ${\cal D}_1$, first made for 
$T=0$ lattices, supports the idea that strong fields are dominantly
selfdual or antiselfdual.

It was only for $T=0$ that the newer data have been analyzed 
by the authors themselves
\cite{DiGiacomo2,DiGiacomo3,DiGiacomo4} . 
As short distances became observable at higher $\beta$   
it has been confirmed that the functions ${\cal D}$ and ${\cal D}_1$ 
contain a perturbative part which could be subtracted by a fit like
(\ref{eq:nonpert_pert}). For the case of quenched $SU(3)$ gauge theory
the two correlation lengths have been found equal, 
$\Lambda_0\approx\Lambda_1\approx0.22$ fm. \footnote{For quenched $SU(2)$
gauge theory, a first analysis \cite{Campostrini} without cooling has found 
$\Lambda_0\approx0.17$ fm.} The data for full QCD show that the correlation 
length $\Lambda_0$ becomes smaller with increasing quark mass towards 
the quenched case. For the chiral limit an upper limit for the correlation
length can be estimated as $\Lambda_0\approx\Lambda_1\approx0.45$ fm.
The gluon condensate $G_2$ as extracted from the correlator
according to (\ref{eq:nonpert_pert},\ref{eq:non_pert}) 
is much larger in pure Yang-Mills theory than any phenomenological value.

In Ref. \cite{IMMMS} the field strength correlator data at intermediate
distance have been used
to extract the two parameters characterizing the random instanton liquid 
model \cite{RILM} which gives a good description of most of the hadronic 
correlators at zero temperature. \footnote{Strictly speaking, with
the classical shape of an instanton in the trivial vacuum 
the asymptotic behavior at $x^2 \rightarrow \infty$ of the correlator 
${\cal D}(x)$ can not be described. The inclusion of the interaction with 
the vacuum medium into the ``classical'' Euclidean field equation results in 
a modified instanton profile far from the center \cite{Diakonov_fremon}.
For a recent interesting attempt to find a modified instanton solution
and to use it for the correlator see Ref \cite{Dorokhov99}.}
The size distribution of this model is approximated as 
$d(\rho)=\delta(\rho-\bar{\rho})$, so density $n_4$ and size $\bar{\rho}$
are the only parameters. The investigation in Ref. \cite{IMMMS} was carefully 
dealing with the path dependence of the Schwinger factor in the instanton 
background. For a straight line path, a reasonable description over the 
measured range of distances has been achieved with parameters roughly 
compatible with phenomenological estimates where this is possible: 
density $n_4=0.5~\mbox{fm}^{-4}$ and size $\bar{\rho}=0.44$ fm for full QCD 
(with four flavors), to compare with $n_4=4~\mbox{fm}^{-4}$ and 
$\bar{\rho}=0.3$ fm  in pure $SU(3)$ Yang-Mills theory. 
Second order terms in $n_4$ give subleading contributions to ${\cal D}$ and  
are the leading ones in ${\cal D}_1$.  As long as deviations from selfduality 
or antiselfduality are entirely due to $II$, $\bar{I}\bar{I}$ and $I\bar{I}$ 
overlap, this analysis has led to the conclusion that ${\cal D}_1 < 0$ is 
characteristic for the instanton liquid at $T=0$. 
Although this was not the case in the original fit given by the authors, it
has been shown that the quenched data \cite{DiGiacomo2,DiGiacomo3} 
are compatible with a small negative ${\cal D}_1$ as well. 
With the second order terms taken into account for pure Yang-Mills theory, 
the best fit leads to a change towards smaller $n_4=3.4~\mbox{fm}^{-4}$ and 
$\bar{\rho}=0.27$ fm. The smallness of the correction {\it a posteriori} 
justifies the dilute gas approach to pure Yang-Mills theory and even more so 
for full QCD.  As a rule one can summarize that instanton radii estimated in 
this way are $20$ \% to $30$ \% bigger than the (exponential) correlation 
length of ${\cal D}$.

\subsection{Finite Temperature}	

At finite temperature, the $O(4)$ space-time symmetry is broken to the
spatial $O(3)$ symmetry and the number of independent correlators
is enlarged\cite{SimonovTne0}.
All of them separately depend on 
$|{\bf x}_1-{\bf x}_2|$ and $(x^4)_1-(x^4)_2$. We consider 
\begin{eqnarray}
\label{eq:electr}
\langle 0|Tr\{E_{i}(x_1) S(x_1,x_2)
E_{j}(x_2) S^{\dagger}(x_1,x_2)\} |0 \rangle =\nonumber \\ 
~~~~~~~~~~~~~\\
\delta_{ij}\left({\cal D}^E +{\cal D}_1^E+x_4^2
\frat{\partial {\cal D}_1^E}{\partial x_4^2}\right)+
x_ix_j\frat{\partial {\cal D}_1^E}{\partial {\vf x}^2}
 ~,\nonumber
\end{eqnarray}
where $E_i=G_{i4}$ is the electric field, and 
\begin{eqnarray}
\label{eq:magnet}
\langle 0|Tr\{B_{i}(x_1) S(x_1,x_2)
B_{j}(x_2) S^{\dagger}(x_1,x_2)\} |0 \rangle =\nonumber \\ 
~~~~~~~~~~~~~\\
\delta_{ij}\left({\cal D}^B +{\cal D}_1^B+{\vf x}^2
\frat{\partial {\cal D}_1^B}{\partial {\vf x}^2}\right)-
x_ix_j\frat{\partial {\cal D}_1^B}{\partial {\vf x}^2}
 ~,\nonumber
\end{eqnarray}
where $B_i=\frac12~\epsilon_{ijk}G_{jk}$ is the magnetic field.
Let us assume that $x=(0,0,|{\vf x}|,0)$ and define longitudinal and transversal 
correlators for electric and magnetic fields
\begin{eqnarray}
\label{eq:survey}
{\cal D}_\parallel^E ={\cal D}^E +{\cal D}_1^E+{\vf x}^2
\frat{\partial {\cal D}_1^E}{\partial {\vf x}^2}
~\ ~\ (i=j=3 {\rm ~~in~~ eq.(\ref{eq:electr}}))\nonumber \\
~~~~~~~~~~~~~\nonumber\\
{\cal D}_\perp^E ={\cal D}^E +{\cal D}_1^E
~\ ~\ (i=j=1,2 {\rm ~~in~~ eq.(\ref{eq:electr}}))\nonumber \\
~~~~~~~~~~~~\\ 
{\cal D}_\perp^B={\cal D}^B +{\cal D}_1^B
~\ ~\ (i=j=3 {\rm ~~in~~ eq.(\ref{eq:magnet}}))\nonumber\\
~~~~~~~~~\nonumber\\
{\cal D}_\parallel^B ={\cal D}^B +{\cal D}_1^B+{\vf x}^2
\frat{\partial {\cal D}_1^B}{\partial {\vf x}^2}
~\ ~\ (i=j=1,2 {\rm ~~in ~~eq.(\ref{eq:magnet}}))~.
\nonumber 
\end{eqnarray}
In addition there is an electric-magnetic correlator which will not be 
considered here. 

As a first step, in Ref. \cite{IMM} an exponential plus perturbative 
fit in terms of ${\cal D}^{E,B}$ and ${\cal D}_1^{E,B}$ has been tried 
to describe the finite $T$ quenched $SU(3)$ correlator measured in 
\cite{DiGiacomo3} with cooling (which is known at distances $> 0.4$ fm).
For the magnetic correlators, immediately below the deconfining transition,
at a temperature $T=0.978~T_{\mathrm{dec}}$,
the two structure functions have clearly different correlation lengths, 
$\Lambda^B_0=0.19$ fm and $\Lambda^B_1=0.53$ fm. The first one slowly increases
across the transition, while the second goes through a maximum there. 
For the two electric structure functions at the temperature 
$T=0.978~T_{\mathrm{dec}}$, different correlation lengths  
$\Lambda^E_0=0.17$ fm and $\Lambda^E_1=0.41$ fm have been found, too.
These fits show a tendency for both electric correlation lengths to drop 
suddenly at the transition.
Unfortunately, there is no cooling data for pure $SU(2)$ at finite temperature
that would allow a more direct comparison.
A description using the Harrington-Shepard caloron solution, i.e., 
the periodic instanton \cite{Harrington,GPY}, is the main objective in 
Ref. \cite{IMM}.
Below the deconfinement temperature, where ${\cal D}^B \approx {\cal D}^E$ 
and ${\cal D}_1^B \approx {\cal D}_1^E$ has been found and where 
${\cal D}^{E,B} >> {\cal D}_1^{E,B}$, 
a satisfactory description of magnetic and electric correlators was possible
using these selfdual or antiselfdual configurations while the small differences
could be attributed to overlapping two-caloron (or caloron-anticaloron) 
configurations. The ratio ${\cal D}_1^E/{\cal D}^E$ strongly increases above 
the transition.

After describing our smoothing method in the next section, we will discuss 
the correlation lengths for the different finite temperature structure 
functions according to exponential Ans\"atze similar to what has been 
done in Ref. \cite{IMM} for the cooled $SU(3)$ data, making combined fits 
to electric and magnetic data separately, in terms of ${\cal D}^E$ and 
${\cal D}_1^E$ and in terms of ${\cal D}^B$ and ${\cal D}_1^B$, respectively.
We will also discuss a fit which directly applies to the {\it raw data} for 
${\cal D}_\parallel^E$ and ${\cal D}_\perp^E$ simultaneously, as well as for
${\cal D}_\parallel^B$ and ${\cal D}_\perp^B$. In this case we will
separate the data into ${\cal D}$ and ${\cal D}_1$ afterwards, using the 
obtained form of the fit. 
It should be emphasized already here that, due to our smoothing method, 
these functions are obtained essentially free of a perturbative contribution. 

\section{Lattice Methods}

\subsection{Smoothing vs. Cooling}

The Pisa group \cite{DiGiacomo1,DiGiacomo2,DiGiacomo3,DiGiacomo4} 
has used the cooling method in order to measure the field strength correlators 
at $T=0$, both without external sources (in vacuum) and in the presence of an
external Wilson loop \cite{DelDebbio}, \footnote{Bali et al. \cite{Bali} have 
analyzed the correlator in the presence of external charges, analyzing quenched
lattice data for heavy quark forces adopting certain factorization assumptions 
from SVM.} and at temperature near $T_{\mathrm{dec}}$. For the measurement, 
similar to the search for instantons, it turned out necessary  to eliminate 
ultraviolet fluctuations. It has been argued that, due to the diffusive nature 
of cooling, the correlator to be measured at a distance $da$ ($a=$ lattice 
spacing) would not be affected before a number $n_{\mathrm{cool}} \propto d^2$ 
of cooling iterations would have been performed. Practically, one measures 
the expectation values defining  
${\cal D}_\parallel(da)$ and ${\cal D}_\perp(da)$ as a function of 
$n_{\mathrm{cool}}$. The actual correlator is then defined as the plateau value
{\it vs.} $n_{\mathrm{cool}}$. Although it has not been presented in detail, 
experience seems to justify the expectation that this plateau is reached 
safely before $|x_1-x_2|$ happens to fall within the diffusion radius
$d \propto \sqrt{n_{\mathrm{cool}}}$. In another context it has been shown 
that this amount of cooling amounts to a renormalization of coupling, 
which is expected to show up in the perturbative signal contributing to the 
correlator.

If cooling is technically necessary to improve the signal-to-noise ratio,
one could be concerned to what extent it distorts the non-perturbative 
modes that existed in the true vacuum,  i.e.,  to what extent the
correlation lengths and the strength of the correlators are affected by
the procedure itself. 
Since the (exponential) correlation length of the correlator ${\cal D}$
is even smaller than the phenomenological instanton size it is useful 
to recall what effect cooling has in the attempt to fix instanton 
size and density from analyzing the topological density of a configuration. 
For this purpose, the cooling technique nowadays is superseded by adopting two,
essentially opposite extreme ways to control that effect. 
The first prescription requires to observe the average size as function of 
$n_{\mathrm{cool}}$, extrapolating it backward to zero cooling steps 
\cite{Boulder_smearing} in order to obtain the would-be instanton size in 
vacuum.
The other uses a global stopping criterion \cite{UKQCD} requiring cooling 
to be stopped when the density of instantons becomes equal to the (much 
better known!) topological susceptibility. This second method implicitly 
adopts a non-interacting instanton gas picture, which differs principally 
from the instanton liquid.  It is recognized by now that one has to control 
cooling since instantons, once they become discernible by cooling or 
smearing, are steadily growing, becoming more and more classical, while 
small size instanton-antiinstanton pairs annihilate. Improved cooling 
\cite{Forcrand} aims to stabilize instantons with $\rho>\rho_{\mathrm{min}}$, 
still suffering from the latter problem. Presently, there is no method 
agreed upon, such that the instanton size cannot be uniquely defined on the 
lattice \cite{Negele}. Each of the methods seems to have a smoothing
radius of the lattice fields (depending on $n_{\mathrm{cool}}$) which
should be well below the distances at which one measures the correlator.
\footnote{Very recently a local stopping criterion for cooling has been 
proposed \cite{Garcia} that seems to accomplish a tunable, well-defined 
smoothing radius.}

During the last two years a method has been advocated \cite{Feurstein,PRD98}
that avoids propagating, iterative procedures to suppress UV fluctuations.
Alternatively, it uses first blocking and subsequent inverse blocking 
with respect to a perfect action in order to accomplish the necessary 
smoothing of equilibrium configurations, while it preserves their long range 
structure beyond the smoothing scale.
So far, this method has been used exclusively to analyze the instanton 
\cite{Feurstein} and monopole structure \cite{Feurstein,PRD98,PRD99} 
of Monte Carlo generated configurations. As for monopoles, smoothing wipes 
out small monopole loops which are irrelevant for confinement (similar to 
blockspin renormalizations directly applied to the magnetic currents 
\cite{Suzuki,Suganuma}).

\subsection{Smoothing with a Perfect Action}

To carry out a theoretically and technically consistent analysis, 
one has to implement a fixed point (FP) action. 
The simplified fixed-point action \cite{Boulder96} for $SU(2)$ has been
improved recently, and a suitably truncated version \cite{Kovacs_private}
has been obtained. It is also parametrized in terms of only two types
of Wilson loops, plaquettes (type $C_{1}$)
\begin{equation}\label{eq:plaq}
U_{C_{1}}=U_{x,\mu,\nu}= 
U_{x,\mu}
U_{x+{\mu},\nu}
U_{x+{\nu},\mu}^+
U_{x,\nu}^+  
\end{equation}
and tilted $3$-dimensional $6$-link loops (type $C_{2}$) of the form
\begin{equation}\label{eq:sixlinks}
U_{C_{2}}=U_{x,\mu,\nu,\lambda}=
U_{x,\mu}
U_{x+{\mu},\nu}
U_{x+{\mu}+\hat{\nu},\lambda}
U_{x+{\nu}+\hat{\lambda},\mu}^+
U_{x+{\lambda},\nu}^+
U_{x,\lambda}^+ \, \, ,
\end{equation}
and contains several powers of the linear action terms
corresponding to each loop (of both types) that can be drawn on the
lattice
\begin{equation}\label{eq:fp_action}
S_{FP}(U)=\sum_{i=1}^{2} \sum_{C_{i}} \sum_{j=1}^{4}
w(i,j) (1 - {1 \over 2}~\mbox{Tr}~U_{C_{i}} )^j  \, \, .
\end{equation}
The parameters of this action as used in this work 
are reproduced in Table I.
A comparison reported in detail in \cite{PRD98} shows that 
the new parametrization is near to perfect.

With this action at hand one can now proceed to perform 
renormalization group transformations on the lattice. 
Suppose we have a fine lattice of links $U$ and a coarse lattice of
links $V$ covering the same physical volume.
Being a classically perfect action, $S_{FP}$ can be evaluated on both
lattices and must satisfy the following
condition
\begin{equation}\label{eq:equality}
~~~~   \mbox{Min}_{U} \left(S_{FP}(U) 
       + \kappa ~ T(U,V)\right) = S_{FP}(V)
\, \, ,
\end{equation}
configuration by configuration over a representative ensemble
of equilibrium gauge field configurations $V$,
each in one-to-one correspondence to a fine configuration $U$ saturating
the lower bound on $S_{FP}(U)+ \kappa ~ T(U,V)$ provided by the right hand side.
$T(U,V)$ is a certain non-negative functional related to the blockspin
transformation (see \cite{Boulder96,Feurstein}). Eq. (\ref{eq:equality})
can be read as the saddle point equation for an integral over fine 
configurations defining the blocked action $S_{FP}(V)$. Neglecting the 
one-loop corrections is justified in the classical limit for 
$\beta \rightarrow \infty$. The parameter $\kappa=12.0$ (not to be confused 
with the non-Abelianicity factor mentioned above) has been fixed in the context
of the adopted blocking scheme by an optimization of the locality properties 
of the quadratic part of the action. 

Blocking means a mapping from the link fields $U$ on the fine lattice to link 
variables $V$ which describe the same configuration on the next coarser length 
scale. This mapping is accomplished by the following construction: 
\begin{eqnarray}\label{eq:blocking}
\tilde{V}_{x,\mu} & = & c_{1}^{block}~U_{x,\mu} U_{x+\hat{\mu},\mu}
\nonumber \\
&+& \sum_{\nu \neq \mu} c_{2}^{block} \left(
U_{x,\nu} U_{x+\hat{\nu},\mu} U_{x+\hat{\nu}+\hat{\mu},\mu}
U_{x+2\hat{\mu},\nu}^{+}  + 
U_{x-\hat{\nu},\nu}^{+} U_{x-\hat{\nu},\mu}
U_{x-\hat{\nu}+\hat{\mu},\mu}
U_{x-\hat{\nu}+2\hat{\mu},\nu} \right) \, \, .   \nonumber
\end{eqnarray}
The coarse link variable $V_{x,\mu}$ is obtained by normalization to a
$SU(2)$ group element 
$V_{x,\mu} 
= {\tilde{V}_{x,\mu}/\sqrt{{\mathrm det}\left(\tilde{V}_{x,\mu}\right)}}$.
The blocking parameters 
$c_{2}^{block} = 0.12$ and
$c_{1}^{block} = 1 - 6~ c_{2}^{block}$ have been optimized earlier in 
Ref. \cite{Bern95} under rather general conditions.

Inverse blocking is  a mapping $V \to U^{SM}$ where the new link variables
$U = U^{SM}$ are implicitly defined as the set of links providing, for the 
configuration encoded in $V$, the ``smoothest interpolation'' on the next finer 
grained lattice. Practically, the link field $U^{SM}$ is found by minimizing 
the extended action on the left hand side of Eq. (\ref{eq:equality}). 
This is not a truly classical field on the lattice but it is smoother than any
quantum field $U$ that could give rise to the coarse level field $V$.
Technically speaking, $\kappa$ plays here a role similar to a Lagrange 
multiplier defining the strength of the constraint imposed by the coarse 
grained configuration.

The $SU(2)$ pure gauge field configurations to be analyzed below have been 
generated on a $12^3 \times 4$ lattice at various $\beta$ with the (classical) 
perfect action.  In Ref. \cite{PRD98} we have localized the deconfinement 
phase transition at $\beta_c=1.545(10)$. Corresponding to the second order 
character of that phase transition, the intersection of Binder cumulants of 
Polyakov loops for different $3d$ lattice volumes had been used for this 
purpose. Although the simulation was done on the fine lattice, the Binder 
cumulants built from $V$ lattice Polyakov loops give the same $\beta_c$. 
While the Polyakov line correlators become critical at the transition, the 
various field strength correlation lengths remain finite across the transition.

\subsection{Measuring the Field Strength Correlators}

Here we demonstrate how we can get immediate access to the non-perturbative 
part of the field strength correlators applying the RG smoothing method. 
With some reservations, the results can be compared with cooling analyses 
(unfortunately available only for $SU(3)$). The field strength components 
$g~a^2~G_{\mu\nu}(x)$ are defined as the  average of the antihermitean part 
$\frac{1}{2~i} \left(U_{x,\mu,\nu}-U_{x,\mu,\nu}^+\right)$
of the four equal-oriented ``clover leaf'' plaquettes in the $\mu\nu$ plane
which are open at the lattice site $x$:
\be
g~a^2~G_{\mu\nu}(x)=\frac{1}{8~i} 
 \left(U_{x,\mu,\nu}+U_{x,\nu,-\mu}+U_{x,-\mu,-\nu}+U_{x,-\nu,\mu} 
 - \mathrm{h.c.} \right) \, .
\label{eq:gemunu}
\ee
For the purpose of the following study we take advantage of the fact that 
in the result of smoothing, similar to cooling, 
the renormalization constant becomes $Z \approx 1$ for the gluon fields.

We have measured only equal-time correlators, {\it i.e.} with a timelike 
argument $(x^4)_1-(x^4)_2=0$.  For lattice sites $x_1,x_2$ 
separated not along one of the spacelike axes, the choice of
the Schwinger line is not unique. In that case 
averaging of expression (\ref{eq:basic_def}) has been performed over a random 
selection of ten shortest paths connecting $x_1$ and $x_2$, 
each path defining an independent transporter $S(x_1,x_2)$.
Naturally, for ${\cal D}_\parallel^E$,  ${\cal D}_\parallel^B$,  
${\cal D}_\perp^E$, and ${\cal D}_\perp^B$, the field components
${\vf E}_\parallel$, ${\vf B}_\parallel$, ${\vf E}_\perp$ and ${\vf B}_\perp$ 
parallel or perpendicular to the 3-vector ${\vf x}_1-{\vf x}_2$
have been appropriately projected out of the field strength tensor
$G_{\mu\nu}$.

\section{Results}

The raw data for $\beta=1.4$ and $\beta=1.5$ (confinement),
for $\beta=1.55$ (just above deconfinement) and three temperatures
in the deconfined phase ($\beta=1.6$, $\beta=1.7$ and $\beta=1.8$)
are shown in Fig. 1, together with an ad hoc fit to be discussed
below. The two correlation functions ${\cal D}_{\perp}^E$ and 
${\cal D}_{\parallel}^B$ have to be understood as averages over two 
correlators involving 
mutually parallel components of ${\vf E}$ or ${\vf B}$ orthogonal to 
${\vf x}_1 - {\vf x}_2$. For better visibility the sum is shown in Fig. 1.
The data shown are averaged over correlators with exactly equal distance.
Even then the data still would show an eventual rotational non-invariance 
(in the three-dimensional distance) if rotational symmetry would be seriously
broken.

\subsection{Two-Exponential Fit}

At first, analogously to the procedure of DiGiacomo et al. applied to their 
$T=0$ cooled data \cite{DiGiacomo2,DiGiacomo3,DiGiacomo4}, 
we have adopted the decomposition (\ref{eq:survey}) into the basic structure 
functions for the ${\vf E}$ and ${\vf B}$ correlators separately,
here however without a perturbative part 
which is suppressed by the steps of blocking and inverse blocking 
(see Sec. IV.B). 
We have assumed the following form
for the basic ${\vf E}$ and ${\vf B}$ correlators 
\bea\label{eq:pure_exp}
{\cal D}({\vf x}) &=& A_0 {\rm e}^{-|{\vf x}|/\Lambda_0} \, ,
\nonumber\\ [-.2cm]
\\ [-.25cm]
{\cal D}_1({\vf x}) &=& A_1 {\rm e}^{-|{\vf x}|/\Lambda_1} ~.
\nonumber
\eea
Only correlator data at distances $d \geq 2~a$ have been
included into the fit, in accordance to our understanding that fields below
that scale are interpolated in a smooth way and taking the extendedness of 
the field strength operator (\ref{eq:gemunu}) into account.
The four corresponding parameters are given in Table II
to allow a comparison with the amplitudes and exponential
correlation lengths obtained for $T\approx T_{\mathrm{dec}}$ 
by the cooling method for $SU(3)$ \cite{DiGiacomo3} as analyzed 
in Ref. \cite{IMM}.

First, we mention that all structure functions are found positive
at these temperatures in both cases. This property is corroborated also 
in the second way of fitting to be discussed below.
Generally, at all temperatures, the two-exponential fit does not properly 
describe the decay of ${\cal D}^B_{\parallel}$ as measured here by the 
smoothing method at the largest distances.  Due to cancellation in
the expression
$${\cal D}^B +{\cal D}_1^B
+{\vf x}^2 \frat{\partial {\cal D}_1^B}{\partial {\vf x}^2}$$
this function would drop too fast. The two correlation lengths associated with
${\cal D}^B$ and ${\cal D}^B_1$ by the fit are almost degenerate and do
not grow with temperature.

For the ${\vf E}$ correlators, a correlated fit of ${\cal D}^E_{\parallel}$
and ${\cal D}^E_{\perp}$ starts to be increasingly inadequate at temperatures 
$T > T_{\mathrm{dec}}$. An indication is the ${\vf E}$ correlator amplitude 
$A_0$ for $T=1.15~T_{\mathrm{dec}}$ where the signal has an uncertainty 
of the same order of magnitude. Taking the parameters of this fit seriously, 
the most dramatic feature is the drop of the correlation length of ${\cal D}^E$
across the phase transition. At $\beta=1.4$ ($T=0.71~T_{\mathrm{dec}}$),
safely below the deconfining transition, we observe a weak splitting of the 
two correlation lengths, with $\Lambda_0\approx 0.22$ fm. At 
$T=0.9~T_{\mathrm{dec}}$ we find still $\Lambda_0\approx 0.18$ fm. 
Immediately above the deconfining transition (at $T=1.02~T_{\mathrm{dec}}$) 
it has dropped already to $0.10$ fm.
 
\subsection{Gauss-Exponential Fit}

For the purpose of the fit shown in Fig. 1, we assumed all four functions 
to have a form like $A~\exp\left(-B~|{\vf x}|-C~|{\vf x}|^2\right)$ 
and required the same limit for $|{\vf x}|\to 0$ for ${\cal D}_\parallel^E$ 
and ${\cal D}_\perp^E$, as well as for ${\cal D}_\parallel^B$ and
${\cal D}_\perp^B$. 

In addition to this, have we considered remaining perturbative contributions 
to these four functions being proportional to $1/|{\vf x}|^4$.
For all temperatures and all correlation functions we found this contribution 
to be smaller than the signal by 
one or two orders of magnitude over the range of distances
from 1 to 6 lattice spacings. Of course, also here only correlator data at 
distances $d \geq 2~a$ have been included into the fit. 
In the following we therefor only discuss parameters which were determined by fits neglecting such a power-like perturbative contribution.

The parameters $A$, $B$, and $C$ are listed 
in Table III in lattice units. In the confinement phase, the coefficient $C$ 
is relatively important only for ${\cal D}_{\parallel}^E$ and 
${\cal D}_{\perp}^B$. While it remains roughly unchanged also at higher 
temperature for the magnetic correlator, it rises dramatically in the electric 
case in order to account for the rapid decay. On the basis of this fit, we 
have determined the structure functions ${\cal D}^E$, ${\cal D}_1^E$, 
${\cal D}^B$ and ${\cal D}_1^B$ which are shown in Fig. 2 for all $\beta$ 
values.

In a first step the ${\cal D}_1^{E,B}$ have been numerically obtained by 
integrating the difference ${\cal D}_\parallel^{E,B}-{\cal D}_\perp^{E,B}$ 
(as represented by the fits) from $0$ to $|{\vf x}|$. The integration 
constants ${\cal D}^{E,B}(0)$ are fixed by the condition that 
${\cal D}_1 \to 0$ for $|{\vf x}| \to \infty$. We find the functions 
${\cal D}_1^{E,B}$ positive. Finally, the ${\cal D}^{E,B}$ have been obtained 
subtracting ${\cal D}_1^{E,B}$ from ${\cal D}_{\perp}^{E,B}$. 

The amplitudes ${\cal D}^E(0)$, ${\cal D}_1^E(0)$, ${\cal D}^B(0)$, 
${\cal D}_1^B(0)$ (which are related to the gluon condensate at high temperature) and 
the integrated correlation lengths of ${\cal D}^E$ and ${\cal D}^B$, 
$\xi_{int}^{E,B}=\int_0^{\infty} d|{\vf x}| {\cal D}^{E,B}({\vf x})
/{\cal D}^{E,B}(0)$, are given in Table IV for the six temperatures.
As long as $T<T_{\mathrm{dec}}$ the functions ${\cal D}_1^{E,B}$
are smaller by roughly a factor four than the corresponding ${\cal D}^{E,B}$.
That means that selfdual or antiselfdual fields are less dominant in this
range of temperatures compared to zero temperature.
The ratio ${\cal D}_1/{\cal D}$ has been found \cite{IMM} considerably smaller
below the deconfinement transition in an analysis of $SU(3)$ correlators
obtained with cooling. Since there are no cooling data for $SU(2)$ so far
in our temperature range, we cannot decide whether this is due to the
other gauge group or whether there is a systematic effect of cooling to
enhance the contribution of locally (anti)selfdual configurations to the 
correlator at higher temperature but still in the confined phase.

The condensates ${\cal D}^E(0)$ and ${\cal D}^B(0)$ only slightly
grow across the transition and start to rapidly grow not before
$T> 1.2 ... 1.5~T_{\mathrm{dec}}$. Up to these temperatures
the ${\cal D}^{E,B}_1(0)$ have increased to a size comparable 
with ${\cal D}^{E,B}(0)$.

The integrated correlation lengths of ${\cal D}^{E,B}$
shrink in a rather continuous way over the temperature range considered
from 0.3 fm (electric and magnetic) at $T=0.7~T_{\mathrm{dec}}$ to 
0.13 fm (magnetic) and 0.07 fm (electric) at $1.9~T_{\mathrm{dec}}$ (Fig. 3).
The smooth change of the four structure functions with increasing
temperature is shown in Fig. 2.

\section{Discussion}

At first, we find our data approximately $O(4)$ rotational invariant also
at $T \ne 0$ as long we consider the confinement phase. The small difference 
between the corresponding electric and magnetic correlators 
(both for ${\cal D}_{\parallel}$ and ${\cal D}_{\perp}$) probably can be 
attributed to overlapping fields of (anti)calorons, within a small density 
expansion. This picture, however, breaks down in the deconfined phase.  
In the confinement phase, the correlator ${\cal D}_{\parallel}^{E}$ 
practically coincides with ${\cal D}_{\parallel}^{B}$ (see Fig. 1 and note 
there the upward shift by a factor two in the latter). At the transition, 
the breakdown of ${\cal D}_{\parallel}^{E}$ (in the correlation length) 
is the most notable result of our analysis.

Comparing the two methods of fitting we find that the condensates 
estimated from the constrained, purely exponential fits in terms of two
structure functions ${\cal D}^E$ and ${\cal D}^E_1$ tend to be bigger than 
the condensates from the second fit. This is because the latter better 
accounts for the real shape of the correlators near zero distance.

The exponential fit also does not properly describe the correlator
${\cal D}^B_{\parallel}$ at large distances. This observation holds 
for all temperatures. Within the deconfined phase, the ${\vf E}$ correlators 
cannot be described at all, by an Ansatz in terms of exponential ${\cal D}^E$ 
and ${\cal D}^E_1$ functions.  Thus, the rapid change of the exponential 
correlation lengths (in particular, of the electric correlation length) at 
the transition merely reflects the difficulty to reach a constrained fit 
within these assumptions. 

The Gauss-exponential fit proves to be much more robust over the whole 
range of temperatures. Without any further assumptions concerning the shape 
of ${\cal D}^{E,B}$ and ${\cal D}_1^{E,B}$, such a fit gives access to these
functions which indeed seem to be poorly described by exponentials.
We have to describe them by integrated correlation lengths which cannot 
be directly compared with the exponential correlation lengths of the
supposedly exponential fits. In general, the integrated correlation lengths 
turn out to be bigger than the exponential ones.

In contrast to the exponential correlation lengths of the constrained 
exponential fits, the integrated correlation lengths associated to 
the correlation functions ${\cal D}^{E,B}$ and ${\cal D}_1^{E,B}$,
as reconstructed from the second fit, change rather smoothly.
As the result of these smooth changes with temperature, the correlation 
function ${\cal D}_{\parallel}^{E}$ drops (in correlation length) 
at the deconfinement transition.

The reconstruction of the absolute size of ${\cal D}$ and ${\cal D}_1$ 
structure functions (condensates) from the fits of ${\cal D}_{\parallel}$ 
and ${\cal D}_{\perp}$ becomes increasingly difficult at higher and higher 
temperature. Below the deconfinement temperature and even near to 
$T_{\mathrm{dec}}$ as a characteristic feature of the data, it is 
well-recognized, that the ${\cal D}_1$ part of the electric and the magnetic 
correlator is considerably smaller than the ${\cal D}$ part.

The dominance of (anti)selfdual fields, however, associated with this 
observation is already weaker in our present results, at temperatures below 
the transition, than it has been found for zero temperature.
It is also weaker than what has been found analyzing cooling data for 
quenched $SU(3)$ near the deconfinement temperature. 
In Table V this statement can be checked for the so-called non-Abelianicity 
of the magnetic correlator. 
It is roughly 80 \% at $T=0.7~T_{\mathrm{dec}}$ and drops to
than 70 \% at $T_{\mathrm{dec}}$. In the deconfined phase 
$\kappa^B$ amounts to approximately 50 \%, with an increasing uncertainty.

The positivity of magnetic and electric ${\cal D}_1$ is a clear feature of 
the data. This applies to the correlators obtained in this work by smoothing 
as well as to the measurements using the cooling method (for quenched
$SU(3)$). Within a caloron 
analysis (which successfully works only below $T_{\mathrm{dec}}$ for the 
$SU(3)$ cooling data) there is no compelling reason to have ${\cal D}_1 < 0$. 
This is unlike the instanton liquid description for $T=0$.
A caloron analysis of the present data will be given elsewhere
\cite{in_preparation}.

The sum of ${\cal D}(0)$ and ${\cal D}_1(0)$ is given by the extrapolation
to zero distance of both ${\cal D}_{\parallel}$ and  ${\cal D}_{\perp}$
for the electric and magnetic correlators.
This quantity is known without big uncertainty and is also presented in 
Table V in physical units for the six temperatures.
The behavior of condensates and correlation lengths is sketched 
graphically across and beyond the deconfinement transition in Fig. 3. 
Quantitatively, the correlation lengths are not very different from the
results of the cooling method, applied on $SU(3)$ configurations. 

The present work dealt with $SU(2)$ pure gauge theory instead of
the phenomenologically more interesting case of $SU(3)$. However it 
has shown an interesting pattern of temperature dependence of 
the magnetic and electric correlation lengths related to the 
different structure functions. 
While the basic structure functions 
${\cal D}$ and ${\cal D}_1$ change smoothly across the phase transition,
the most dramatic effect, the breakdown of ${\cal D}_{\parallel}^E$,
results from the interplay. 
A simple picture according to which ${\cal D}^E$ would turn to zero in the
deconfined phase is not corroborated by our data. 

It seems to be worthwhile to investigate the field strength correlators for
$ SU(2)$ using the cooling method, too, in order to compare it with the
smoothing method applied here.
It would be even more interesting to implement the RG smoothing method 
for the gauge group $SU(3)$. This would then allow for a 
direct comparison with the pioneering results already available thanks 
to the Pisa group.

\begin{table}[h]
\begin{center}
\begin{tabular}{ l c c c c }
$w(i,j)$            & $j=1$   &  $j=2$  & $j=3$ & $j=4$   \\
\hline
$i=1$ (plaquettes)  &    $1.115504$& $-0.5424815$ & $0.1845878$     & $-0.01197482$ \\
$i=2$ (6-link loops)& $-0.01443798$& $0.1386238 $ & $-0.07551325 $ & $0.01579434$ 
\\
\end{tabular}
\end{center}
\caption{ Weight coefficients of the simplified fixed-point action for $SU(2)$. }
\label{tab:weights}
\end{table}

\begin{table}[t]
\label{tab:expfit}
\vspace{0.2cm}
\begin{center}
\footnotesize
\begin{tabular}{ l l l c c c c c }
Type & {$\beta$} &  {$T/T_{\mathrm{dec}}$} & $A_0/T_{\mathrm{dec}}^{4}$ & $\Lambda_0\times T_{\mathrm{dec}}$ & $A_1/T_{\mathrm{dec}}^{4}$ & $\Lambda_1\times T_{\mathrm{dec}}$ & $\chi^2$/d.o.f. \\
\hline
$E$ &1.40 &0.71 &$3.92(2)$     &$0.326(5)$   &$1.59(1)  $   &$0.354(1)$& 15.92 \\
$E$ &1.50 &0.90 &$6.94(5)$     &$0.274(6)$   &$3.09(3)  $   &$0.314(1)$& 10.11 \\
$E$ &1.55 &1.02 &$17.4(8)$     &$0.156(2)$   &$5.016(6) $   &$0.318(1)$& 10.84 \\
$E$ &1.60 &1.15 &$270(290)$    &$0.055(6)$   &$12.80(4) $   &$0.2520(3)$&27.33 \\
\hline
$B$ &1.40 &0.71 &$3.50(1)$     &$0.3455(5)$  &$1.45(1)  $ &$0.368(1)$ & 7.271\\
$B$ &1.50 &0.90 &$6.59(3)$     &$0.2920(6)$  &$3.04(3)  $ &$0.310(1)$ & 4.526\\
$B$ &1.55 &1.02 &$7.03(6)$     &$0.2832(7)$  &$4.45(3)  $ &$0.290(1)$ & 0.823\\
$B$ &1.60 &1.15 &$7.41(4)$     &$0.2760(8)$  &$6.51(4)  $ &$0.273(1)$ & 0.294\\
$B$ &1.70 &1.47 &$12.1(1)$     &$0.2386(9)$  &$14.9(1)  $ &$0.2272(8)$& 0.448\\
$B$ &1.80 &1.88 &$23.2(3)$     &$0.2004(8)$  &$35.2(3)  $ &$0.1853(7)$& 0.660\\
\end{tabular}
\end{center}
\caption{Condensates and correlation lengths obtained from exponential 
fits of the form
${\cal D}^{E,B}({\vf x}) = A_0~\exp \left(-|{\vf x}|/\Lambda_0\right)$ 
and 
${\cal D}_1^{E,B}({\vf x}) = A_1~\exp \left(-|{\vf x}|/\Lambda_1\right)$
for various temperatures $T$.  The data was fitted for lattice distances 
$d > 2a$. Type $E$ refers to ${\cal D}^E$ and ${\cal D}_1^E$, type $B$ 
to ${\cal D}^B$ and ${\cal D}_1^B$. 
} 
\end{table}

\begin{table}[t]
\label{tab:gaussfit}
\vspace{0.2cm}
\begin{center}
\footnotesize
\begin{tabular}{ l l c c c c }
Type & {$\beta$} &   A  & B  &  C  & $\chi^2$/d.o.f. \\
\hline
${\cal D}^E_{||   }$ & 1.40 &  0.039(3)   & 0.628(2)  &  0.1372(7)&0.278 \\
${\cal D}^E_{\perp}$ & 1.40 &  0.039(1)   & 0.6314(7) &  0.0503(2)&0.278 \\
${\cal D}^B_{||   }$ & 1.40 &  0.038(1)   & 0.5724(6) &  0.0599(2)&0.245 \\
${\cal D}^B_{\perp}$ & 1.40 &  0.038(7)   & 0.570(1)  &  0.1382(5)&0.245 \\
\hline
${\cal D}^E_{||   }$ & 1.50 &  0.030(3)   & 0.584(3)  &  0.137(1) &0.137 \\
${\cal D}^E_{\perp}$ & 1.50 &  0.030(1)   & 0.5992(9) &  0.0444(2)&0.137 \\
${\cal D}^B_{||   }$ & 1.50 &  0.029(1)   & 0.5178(7) &  0.0584(2)&0.162 \\
${\cal D}^B_{\perp}$ & 1.50 &  0.029(6)   & 0.525(2)  &  0.1306(6)&0.162 \\
\hline
${\cal D}^E_{||   }$ & 1.55 &  0.022(8)   & 0.48(1)   &  0.255(5) & 0.044\\
${\cal D}^E_{\perp}$ & 1.55 &  0.022(2)   & 0.623(01) &  0.0369(3)& 0.044\\
${\cal D}^B_{||   }$ & 1.55 &  0.023(1)   & 0.4972(8) &  0.0509(2)& 0.070\\
${\cal D}^B_{\perp}$ & 1.55 &  0.023(3)   & 0.517(2)  &  0.1269(7)& 0.070\\
\hline
${\cal D}^E_{||   }$ & 1.60 &  0.015(5)   & 0.60(1)   &  0.261(9) & 0.083\\
${\cal D}^E_{\perp}$ & 1.60 &  0.015(2)   & 0.613(1)  &  0.0379(3)& 0.083\\
${\cal D}^B_{||   }$ & 1.60 &  0.019(1)   & 0.499(7)  &  0.0406(2)& 0.051\\
${\cal D}^B_{\perp}$ & 1.60 &  0.019(5)   & 0.616(2)  &  0.0869(6)& 0.051\\
\hline
${\cal D}^E_{||   }$ & 1.70 &  0.012(3)   & 0.46(1)   &  0.411(7) & 0.116\\
${\cal D}^E_{\perp}$ & 1.70 &  0.012(1)   & 0.634(1)  &  0.0306(3)& 0.116\\
${\cal D}^B_{||   }$ & 1.70 &  0.015(1)   & 0.4902(8) &  0.0328(2)& 0.042\\
${\cal D}^B_{\perp}$ & 1.70 &  0.015(1)   & 0.510(2)  &  0.1305(9)& 0.042\\
\hline
${\cal D}^E_{||   }$ & 1.80 &  0.095(7)   & 0.35(1)   &  0.4867(7)& 0.130\\
${\cal D}^E_{\perp}$ & 1.80 &  0.095(1)   & 0.6212(9) &  0.0291(2)& 0.130\\
${\cal D}^B_{||   }$ & 1.80 &  0.013(2)   & 0.5002(7) &  0.0258(2)& 0.043\\
${\cal D}^B_{\perp}$ & 1.80 &  0.013(4)   & 0.539(2)  &  0.1227(8)& 0.043\\
\end{tabular}
\end{center}
\caption{Parameters from the fit to the raw data of the correlators
shown in Fig.~1 in lattice units.  
The parameters $A$, $B$ and $C$ correspond to a respective Ansatz 
for ${\cal D}_{\parallel}^{E,B}$ and ${\cal D}_{\perp}^{E,B}$,  
${\cal D}({\vf x}) = A~\exp \left(-B~|{\vf x}|-C~|{\vf x}|^2\right)$. 
} 
\end{table}

\begin{table}[t]
\label{tab:cond1}
\vspace{0.1cm}
\begin{center}
\footnotesize
\begin{tabular}{ l c c c c c c c }
$\beta$ &$T/T_{\mathrm{dec}}$ &  ${\cal D}^E(0)/T^4_{\mathrm{dec}}$ &${\cal D}^B(0)/T^4_{\mathrm{dec}}$ &$ {\cal D}_1^E(0)/T^4_{\mathrm{dec}}$ &$ {\cal D}_1^B(0)/ T^4_{\mathrm{dec}}$ &$\xi_{int}^B\times T_{\mathrm{dec}}$  &$\xi_{int}^E\times T_{\mathrm{dec}}$ \\
\hline
1.4 &0.71 &$2.00(5) $&$ 2.0(1) $ &$0.54(6) $&$ 0.52(3) $&$0.44(3)  $&$0.42(2) $ \\ 
1.5 &0.90 &$3.9(2)  $&$ 3.7(3) $ &$1.1(1)  $&$ 1.2(3)  $&$0.38(1)  $&$0.34(2) $ \\ 
1.55&1.02 &$4.9(4)  $&$ 4.4(2) $ &$1.1(6)  $&$ 2.0(2)  $&$0.35(2)  $&$0.25(3) $ \\ 
1.6 &1.15 &$3.6(4)  $&$ 4.4(6) $ &$3.1(4)  $&$ 4.2(7)  $&$0.27(4)  $&$0.24(4) $ \\ 
1.7 &1.47 &$8.4(7)  $&$ 8.1(16)$ &$5.2(8)  $&$ 5.5(23) $&$0.25(3)  $&$0.15(3) $ \\ 
1.8 &1.88 &$20.1(57)$&$22.0(20)$ &$9.9(59) $&$19.0(24) $&$0.20(2)  $&$0.10(3) $ \\ 
\end{tabular}
\end{center}
\caption{Condensates and integrated correlation lengths of ${\cal D}^{E,B}$
for the fit presented in Table III for various temperatures $T$.  
All quantities are expressed in comparison to $T_{\mathrm{dec}}$.  }
\end{table}

\begin{table}[t]
\label{tab:cond2}
\vspace{0.1cm}
\begin{center}
\footnotesize
\begin{tabular}{ l c c c c c c }
$\beta$ & ${\cal D}^E(0)+{\cal D}_1^E(0)$ &${\cal D}^B(0)+{\cal D}_1^B(0)$ &  
$\kappa ^B$ &  $\sigma^B$\\
 & [GeV$^4$] & [GeV$^4$]  &    & [GeV$^2$]  \\
\hline
1.4  & 0.020(1)  & 0.020(1)   &  0.79(3)  &  0.10(1) \\
1.5  & 0.039(2)  & 0.039(4)   &  0.75(5)  &  0.14(3) \\
1.55 & 0.047(8)  & 0.051(3)   &  0.69(15) &  0.14(3) \\
1.6  & 0.053(7)  & 0.068(10)  &  0.51(41) &  0.15(5) \\
1.7  & 0.107(12) & 0.107(30)  &  0.60(23) &  0.15(10)  \\
1.8  & 0.237(92) & 0.323(34)  &  0.54(64) &  0.27(19)  \\

\end{tabular}
\end{center}
\caption{Some physically interesting quantities (for $SU(2)$)
in physical units.  The magnetic string tension $\sigma^B$ is 
given by (\ref{eq:sigma}) in terms of ${\cal D}^B$.
For definiteness, we took $T_{\mathrm{dec}}=0.298$ GeV for pure $SU(2)$
gauge theory.}
\end{table}
\newpage

\begin{figure}
\label{rhodaten}
\epsfxsize=15.0cm\epsffile{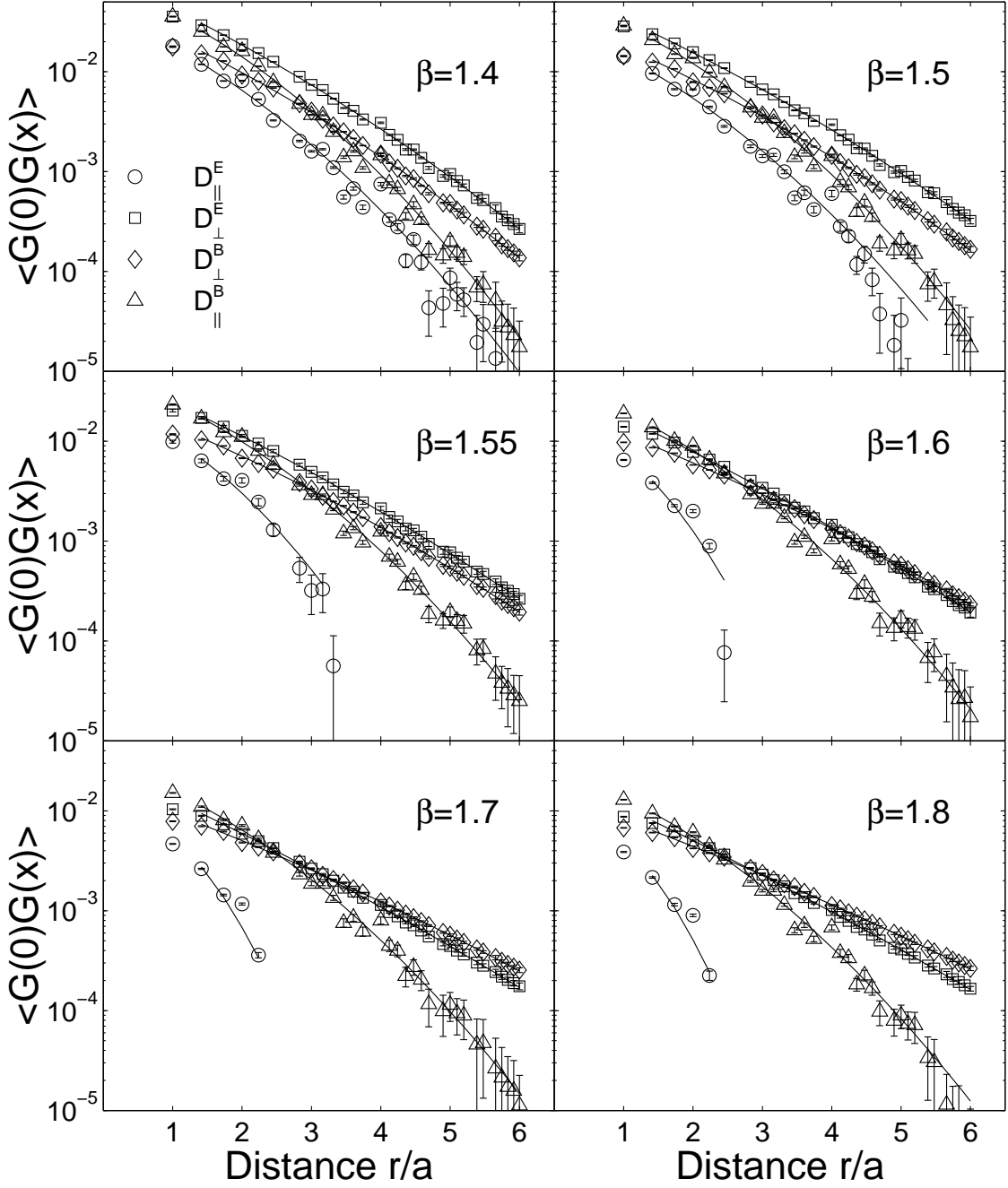} 
\caption{Raw correlator data fitted according to a generic form 
$A~\exp \left(-B|{\bf x}|-C|{\bf x}|^2\right)$. 
The corresponding fit parameters are reported in Table III. 
Errors are determined by the jackknife method for a block size of 
100. The dependence on the block size was found to be marginal.
For the clarity of the plot the data and curves shown for  
${\cal D}^E_{\perp}$ and ${\cal D}^B_{\parallel}$ include a factor two
shift upward.  }	
\end{figure}

\begin{figure}
\label{reconst}
\epsfxsize=15.0cm\epsffile{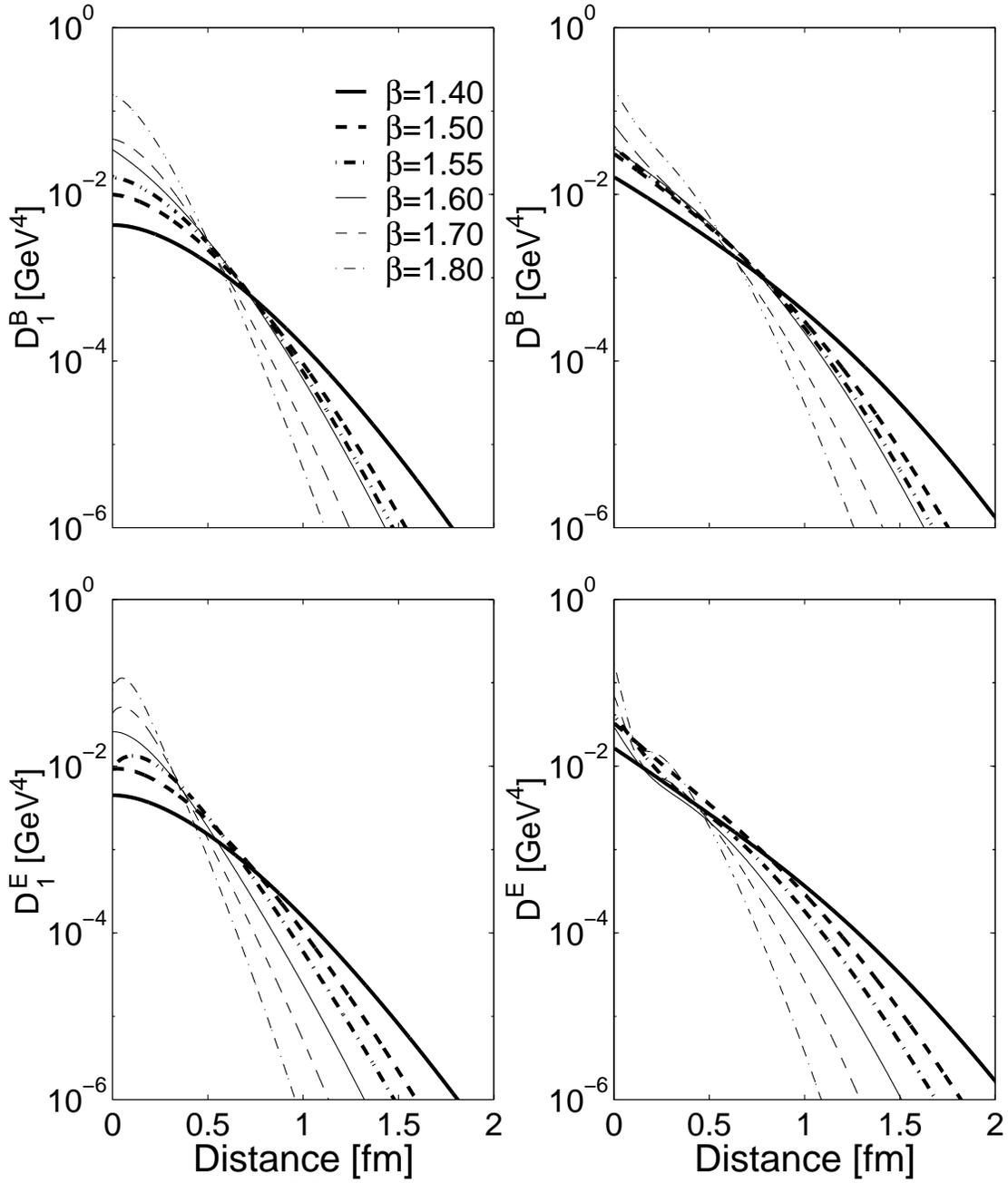} 
\caption{${\cal D}^{E,B}$ and ${\cal D}^{E,B}_1$ reconstructed
from the Gauss-exponential fit of Fig. 1 and Table III, 
here presented in physical units. }
\end{figure}
\begin{figure}
\label{kondens}
\epsfxsize=15.0cm\epsffile{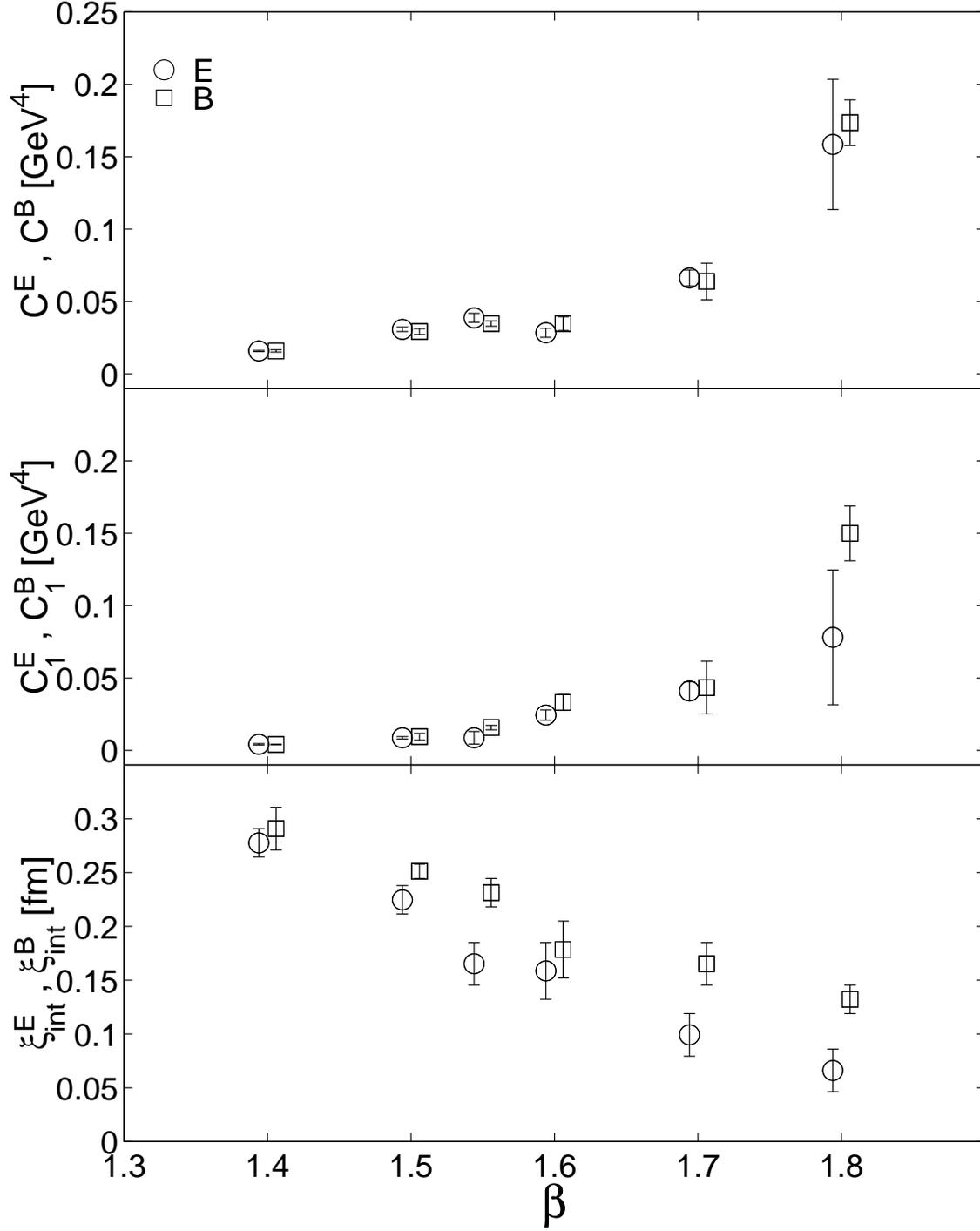} 
\caption{ Condensates and correlation lengths as listed in Table IV, 
here expressed in physical units and shown {\it vs.} $\beta$.  }	
\end{figure}


\end{document}